\documentclass[11pt]{article}
\usepackage{graphicx,amsfonts}
\newtheorem{thm}{T{\sc HEOREM}}[section]
\newtheorem{lemma}[thm]{L{\sc EMMA}}
\newtheorem{prop}[thm]{P{\sc ROPOSITION}}

\def\be{\begin{eqnarray}}
\def\ee{\end{eqnarray}}
\def\ben{\begin{eqnarray*}}
\def\een{\end{eqnarray*}}
\def\ba{\begin{array}}
\def\ea{\end{array}}
\def\bp{\noindent{\it Proof. }}
\def\ep{\noindent{\hfill \fbox{}}}

\def\ul{\underline}
\def\remark{\noindent{\bf Remark. }}
\def\definition{\noindent{\bf Definition. }}
\def\pic{{\rm Pic}}
\def\div{{\rm Div}}

\def\ni{\noindent}
\def\vp{\varphi}
\def\pone{{\mathbb P}^1}
\def\ptwo{{\mathbb P}^2}
\def\ponet{{\mathbb P}^1\times{\mathbb P}^1}
\def\mc{{\mathbb C}}
\def\mz{{\mathbb Z}}
\def\calc{{\cal C}}
\def\cali{{\cal I}}
\def\nn{\nonumber}
\def\ds{\displaystyle}
\newcommand{\ol}[1]{\overline{#1}}
\newcommand{\mapright}[1]{%
   \smash{\mathop{%
   \hbox to 1cm{\rightarrowfill}}\limits^{#1}}}
\newcommand{\mapleft}[1]{%
   \smash{\mathop{%
   \hbox to 1cm{\leftarrowfill}}\limits^{#1}}}
\newcommand{\maplleft}[2]{%
   \smash{\mathop{%
   \hbox to 1cm{\leftarrowfill}}\limits_{#1}^{#2}}}

\begin{document}

\title{The space of initial conditions for linearisable mappings}
\author{T. Takenawa$^1$, M. Eguchi$^1$, B. Grammaticos$^2$, 
Y. Ohta$^3$,\\ A. Ramani$^4$ and J. Satsuma$^1$}
\date{}
\maketitle

\noindent 
1: Graduate School of Mathematical Sciences, University of
Tokyo,\\ \quad Komaba 3-8-1, Meguro-ku, Tokyo 153-8914, Japan\\
2: GMPIB, Universit\'e Paris VII, Tour 24-14, 5$^e$\'etage, 
case 7021, 75251 Paris, France\\
3: Department of Applied Mathematics, Faculty of Engineering, 
 Hiroshima University,\\ \quad
Kagamiyama 1-4-1 , Higashi-Hiroshima, 739-8527  Japan\\
4: CPT, Ecole Polytechnique, CNRS, UMR 7644, 91128 Palaiseau, 
France

\begin{center}
\end{center}
\begin{abstract}
We apply the algebraic-geometric techniques developed for the study of
mappings
which have the singularity confinement property to mappings which are
integrable
through linearisation. The main difference with respect to the previous
studies is
that the linearisable mappings have generically unconfined singularities.
Despite
this fact we are able to provide a complete description of the dynamics of
these
mappings and derive rigorously their growth properties.
\end{abstract}

\section{Introduction}\label{intro}

The study of integrable discrete systems has been pursued with particular
intensity
over the last decade. The results obtained surpassed expectations: not only
does
this reputably difficult domain yield a plethora of new and interesting
results but
it has also been possible to map in detail the parallels and divergences
that exist
between integrable discrete systems and their continuous counterparts. As
paradigms
of integrable discrete systems which were derived and exhaustibly studied
we must
mention here the discrete Painlev\'e equations, i.e. nonautonomous   integrable
mappings, the continuous limit of which are the well-known Painlev\'e
equations.

The progress in the domain of discrete integrability was made possible by the
development of adequate integrability detectors. The first such detector
that has
been proposed  was singularity confinement. It was based on the observation
that for
integrable mappings any spontaneously appearing singularity disappears
after a few
iteration steps. Although singularity confinement was instrumental in the
derivation of several new integrable discrete systems, as for example the
discrete
Painlev\'e equations\cite{rgh}, it became clear that this property in fact
characterises only a restricted class of mappings. The conjecture which we have
proposed in \cite{grp} can be formulated as follows: mappings integrable
through spectral methods have the singularity confinement property. Singularity
confinement in itself is not sufficient in order to control the complexity
of the
iterates of a given mapping. This latter property turned out to be crucial for
integrability and thus new integrability detectors were proposed based on this
feature. Viallet and collaborators \cite{hv,bv} have introduced the notion of
algebraic entropy, which is linked to Arnord's complexity \cite{gizatullin, arnord, cantat}. 
It is defined as $S=\lim_{n\to \infty} \log(d_n)/n$ where
$d_n$ is the degree of the numerator (or denominator)  of the $n$-th iterate of
rational mapping. While nonintegrable mappings exhibit exponential growth
and thus
nonzero algebraic entropy, integrable ones have zero algebraic entropy, their
degree growth being polynomial.

While the detection of integrability is essential, it is by far not
sufficient. As
Kruskal has always kept pointing out, once an integrable mapping is
obtained, one
must still perform the complete study of its dynamics. The difficulty lies
in the
existence of singularities which lead to indeterminate points, i.e. points
where
the iterates of the mapping is not well defined. Fortunately, for mappings
which
have the singularity confinement property, only a finite number of such
points do
exist. In order to deal with these indeterminacies, one must introduce the
adequate
(local) description of the singularities.

A first step towards this  description of integrable discrete systems  was
undertaken by Sakai \cite{sakai} who presented a geometric approach to the
theory of Painlev\'e equations based on rational surfaces, which are called
spaces of initial conditions \cite{okamoto}.
The result was a classification of discrete Painlev\'e equations in terms
of type of rational surfaces and affine Weyl groups.

Further developments in this direction were those introduced by one of
the authors (TT), based on the characterisation of mappings from the theory of
rational surfaces.  In a series of previous works
\cite{takenawa1,takenawa2,takenawa3}
this approach was used for:

a) mappings which are nonintegrable while having the singularity confinement
property. It was possible in that case to obtain rigorously the value of their
algebraic entropy.

b) the mappings of the Sakai classification, for which it was shown that
the degree
growth is up to quadratic.

Here the same approach will be applied to a different class of integrable
mappings:
linearisable ones \cite{rglo,rog}. By  linearisable we mean mappings which can be reduced to a
local linear difference system. The difficulty in the present case lies in
the fact
that, generically, linearisable mappings have {\sl unconfined}
singularities. Thus
one must perform the local study around an infinite number of
singularities. Still,
due to the particular structure of these systems, this turns out to be
possible and
allows us to compute precisely their
degree growth (which coincides with the one previously obtained empirically).
Moreover, the corresponding surfaces provide a straight-forward way
to construct transformations of these mappings to linear mappings in
cascade. 

\section{Preliminaries}\label{pre}

This section is devoted to a review of some basic notions, which  we will
use later,
and  the arrangement of notations. In this paper we consider mainly birational
mappings, but in order  to compare the notions of space of initial
conditions and
analytical stability \cite{sibony,df} 
we consider dominant rational mappings (i.e. meromorphic
mappings such that the closure of each image in the  target manifold is the
manifold itself) instead of birational mappings in this section.\\

\definition (degree of rational mapping)\\
\noindent i) {\it $\ptwo (\mc)$ case.}  Let $\vp_i$: $(x,y)\in \mc^2 \mapsto
(\ol{x},\ol{y})\in \mc^2$  be a rational mapping for $i=0,1,2,\cdots$. We can
relate a mapping
$\vp_i'$: $(X:Y:Z)\in \ptwo \mapsto
(\ol{X}:\ol{Y}:\ol{Z})=(f_i(X,Y,Z):g_i(X,Y,Z):h_i(X,Y,Z)) \in \ptwo$  to the
mapping $\vp_i$ by using the relations $x=X/Z,y=Y/Z$ and $\ol{x}=\ol{X}/\ol{Z},
\ol{y}=\ol{Y}/\ol{Z}$  and by reducing to a common denominator. The degree
of the
sequence of mappings is defined by the degree of polynomials.\\
\noindent ii) {\it ${\mathbb P}^1(\mc) \times {\mathbb P}^1(\mc)$ case.}
We denote
the degree of a polynomial on ${\mathbb C}:$
$f(t)=\sum_{m}a_{t}t^m$ by $\deg f(t)$ $(=\deg_tf(t))$. The degree of a
rational
(=meromorphic) function on ${\mathbb P}^1$,  which is written as
$P(x)=f(x)/g(x)$ on  one of the local coordinates, where
$f(x)$ and $g(x)$ are polynomials,  is defined by
$$\deg(P)=\max\{\deg f(x),\deg g(x)\}.$$ The degree of an irreducible rational
function on ${\mathbb P}^1\times {\mathbb P}^1$,  which is written as
$P(x,y)=f(x,y)/g(x,y)$ on  one of the local coordinates, where
$f(x,y)$ and $g(x,y)$ are polynomials,  is defined by
$$\deg(P)= \deg_x P(x,y)+\deg_y P(x,y).$$ The degree of a mapping $\vp:{\mathbb
P}^1\times {\mathbb P}^1
\to {\mathbb P}^1\times {\mathbb P}^1$, $(x,y)\mapsto(P(x,y),Q(x,y))$, where
$P(x,y)$ and $Q(x,y)$ are rational functions, is defined by
$$\deg(\vp)=\max\{\deg P(x,y),\deg Q(x,y)\}.$$\\

\noindent{\it Blowing up}\\ Let $X$ be a smooth projective surface and let
$p$ be a
point on $X$ . There exists a smooth projective surface $Z$ and a morphism
$\pi:Z
\to X$ such that $\pi^{-1}(p)\simeq \pone$ and $\pi$ is a biholomorphic
mapping from
$Z\setminus \pi^{-1}(p)$ to $ X\setminus p$. The morphism $\pi$ is  called
blow-down and the correspondence $\pi^{-1}$  is called blow-up of $X$ at
$p$ as a
rational mapping.\\

Let $X$ and $Y$ be smooth projective surfaces and let $X'$ and $Y'$ be surfaces
obtained by successive blow-ups
$\pi_X^{-1}: X \to X'$ and $\pi_Y^{-1}: Y \to Y'$. A rational mapping
$\vp': X'\to
Y'$ is called {\it lifted} from a rational mapping $\vp:X\to Y$ if
$\pi_Y \circ \vp'(x')= \vp \circ \pi_X (x')$ holds if 
$\vp'(x')$ and $\vp \circ \pi_X (x')$ are defined.\\

\definition (indeterminate set and critical set)  Let $X$ and $Y$ be smooth
projective surfaces and  let  $\vp: X \to Y$ be a dominant rational
mapping.  The
set of indeterminate points $\cali(\vp)$ is defined by
\be \cali(\vp):=\{ x \in X; \vp \mbox{ is indeterminate at } x \}.
\ee

Let $\pi^{-1}: X \to X'$ be a blowing up that eliminates the indeterminacy
of $\vp$
and let $\vp'$  be the mapping lifted from $\vp$  (hence both $\pi$ and
$\vp'$ are
holomorphic).     The set of ``critical'' points $\calc(\vp)$ is defined by
\be \calc(\vp):=\{ y \in Y;~ \dim(\pi(\vp'^{-1}(y)))\geq 1 \},
\ee 
where $\vp^{-1}(y)$ is considered to be a set correspondence.\\

\remark\\ i) The critical set $\calc(\vp)$ is independent of the choice of
$X'$.\\ ii)
If $\vp$ is birational, then $\calc(\vp) = \cali(\vp^{-1})$.\\

\definition (space of initial conditions) Let $Y_i$  be smooth projective
surfaces and
let $\{\vp_i:Y_i\to Y_{i+1}\}$ be a sequence of dominant rational mappings.  
A sequence of
rational surfaces $\{X_i\}$ is (or $X_i$ themselves are) called  the space  of
initial conditions for the sequence of $\vp_i$ if  each $\vp_i$ is lifted
to the
mapping $\vp_i:X_i\to X_{i+1}$ such that
${\cal I}(\vp_i)={\cal C}(\vp_i)=\emptyset$ (an isomorphism, i.e.
bi-holomorphic mapping
if $\vp_i$'s are birational)
\cite{sakai,okamoto,takenawa3}.\\

\noindent{\it Picard group}\\ We denote the group of divisors on a smooth
projective surface
$X$ as $\div(X)$. The Picard group of $X$ is the group  of isomorphism
classes of
invertible sheaves on $X$ and it is isomorphic to the group of linear
equivalence
classes of divisors on $X$. We denote it by $\pic(X)$.\\

\definition (pull-back and push-forward action for a surjective morphism)
Let  $\vp:
X \to Y$ be a surjective morphism, where $X$ and $Y$ are smooth projective
surfaces.
The pull-back action $\vp^*: \div(Y) \to \div(X)$ is defined naturally as
on the
Cartier divisors.

Let $C$ be an irreducible curve (possibly singular) on $X$.  The push forward
action $\vp_*: \div(X) \to \div(Y)$ is defined by
\be
\vp_*(C)&:=& \left\{
\ba{ll} 0& (\mbox{if $f(C)$ is a point})\\
\lambda \gamma &(\mbox{$\gamma=f(C)$ is an irreducible curve on$Y$}),
\ea \right. \ee where $\lambda$ is the degree of the covering of $\vp|_C: C\to
\gamma$,  and its linear combinations.\\

\remark Since the action $\vp^*$  (or $\vp_*$) maps the principal divisor
on $Y$
 to that of $X$ (resp. that of $X$ to that of $Y$),  it reduces the
homomorphism
from
$\pic(Y)$ to $\pic(X) $ (resp. from $\pic(X)$ to $\pic(Y)$)  (see Chap. 1
\cite{beauville} or Prop.1.4 \cite{fulton}). \\

\definition (pull-back and push-forward action for a rational mapping)  Let
$X$ and
$Y$ be smooth projective surfaces and  let  $\vp: X \to Y$ be a dominant
rational
mapping.  Let $X'$ be a surface obtained by successive blow-ups
$\pi^{-1}: X \to X'$ that eliminates the indeterminacy of $\vp$ and let
$\vp'$  be the
mapping lifted from $\vp$.  The pull back action $\vp^*: \pic(Y) \to
\pic(X)$ is
defined by
$$\vp^*:=\pi_* \circ \vp'^*: \div(Y) \to \div(X)$$ and the push-forward
action is
defined by
$\vp_*$ ( $=(\vp^{-1})^* $ in the case where $\vp$ is birational)
$$\vp_*:= \vp'_* \circ \pi^* : \div(X) \to \div(Y).$$

\remark These actions are well-defined with respect to choice of $X'$.\\

\bp Let $\vp: X \to Y$ be a dominant rational mapping and let $X_1$ and
$X_2$ be
surfaces obtained by eliminating the indeterminacy of $\vp$.  We have the
holomorphic
mappings
$\pi_i:X_i\to X$ and $\vp_i: X_i \to Y$, where $\vp_i$'s  are mapping
lifted from
$\vp$. Moreover there exists a birational mapping $\psi: X_1 \to X_2$ such that
$\pi_1=\pi_2 \circ \psi$. Hence there exists a smooth projective surface
$\tilde{X}$ and morphisms $\tilde{\pi_1}:\tilde{X}\to X_1$,
$\tilde{\pi_2}: \tilde{X}\to X_2$ such that
$\tilde{\pi_2}=\psi \circ \tilde{\pi_1}$. Let $D$ be a divisor on $Y$. We have
\ben
\pi_{1*}\circ \vp_1^* (D) &=& \pi_{1*}\circ \tilde{\pi}_{1*}\circ
\tilde{\pi}_1^*\circ \vp_1^* (D)\\ &=& (\pi_1\circ \tilde{\pi}_1)_* \circ (
\vp_1\circ \tilde{\pi}_1)^*(D)\\ &=& (\pi_2\circ \tilde{\pi}_2)_* \circ (
\vp_2\circ \tilde{\pi}_2)^*(D)\\ &=& \pi_{2*}\circ \tilde{\pi}_{2*}\circ
\tilde{\pi}_2^*\circ \vp_2^* (D)\\ &=& \pi_{2*}\circ \vp_2^* (D),
\een where we use the fact that for any composition of blow-ups
$\pi^{-1}: Z_1 \to Z_2$  and any divisor 
$D \in \div(Z_1)$, $\pi_* \circ \pi^*(D)=(D)$
holds.
\ep\\

\begin{prop} \label{pr1} Let $X,Y$ and $Z$ be smooth projective surfaces
and let
$f:X\to Y$ and $g:Y \to Z$ be a dominant rational mappings. Then for any
effective
divisor $D$ on $Z$
$$ (g\circ f)^* (D) \leq f^*\circ g^* (D)$$   holds, where the equality
holds if
and only if
$\calc(f)\cap \cali(g)=\emptyset$.
\end{prop}

In our case (smooth projective surfaces' case) this proposition can be  proved
simply as follows.\\

\bp

The ``only if'' part is obvious. Let $D$ be an irreducible curve such that
$g^{-1}(D)\subset \calc(f)\cap \cali(g)$.  Then we have $(g\circ f)^* (D) >
0$ and
$f^*\circ g^* (D)=f^*(0)= 0$.

Let us prove the ``if'' part. Without loss of generality we can assume $D$
is an
irreducible curve. There exist irreducible curves $D_1,D_2,\cdots,D_m$, points
$p_1,p_2,\cdots, p_l \subset \cali(g)$ and positive integers
$r_1,r_2,\cdots,r_m$ such that $g^{-1}(D)=D_1 + \cdots + D_m + p_1+ \cdots
+p_l$ and
$g^*(D)=r_1 D_1 +r_2 D_2 +\cdots +r_mD_m$. Similarly for each $D_i$ there exist
irreducible curves
$D_{i,1},D_{i,2},\cdots,D_{i,m_i}$, points
$p_{i,1},p_{i,2},\cdots, p_{i,l_i} \subset \cali(f)$ and positive integers
$r_{i,1},r_{i,2},\cdots,r_{i,m_i}$ such that $g^{-1}(D_i)=D_{i,1} + \cdots +
D_{i,m_i} + p_{i,1}+ \cdots + p_{i,l_i}$ and $g^*(D)=r_{i,1} D_{i,1} + r_{i,2}
D_{i,2} +\cdots + r_{i,m_i} D_{i,m_i}$. Hence we have
$$f^*\circ g^* (D)=\sum_{i=1}^m r_i \sum_{j=1}^{m_i} p_{i,j}D_{i,j}.$$
Since $p_i
\not\in \calc(f)$ and $g|_{D_i}:D_i\to D$ is holomorphic,
$(g\circ f)^* (D)$ is the same quantity.
\ep \\

\definition (analytical stability \cite{sibony}) Let each $X_i$ be a smooth
surface and let
$\{\vp_i:X_i\to X_{i+1} \}_{i=0,1,2,\cdots}$  be a  sequence of dominant
rational
mappings. For a subset $S \subset X_i$ we denote 
$\vp_i(S\setminus \cali(\vp_i))$ as $\vp_i(S)$. 
The sequence $\{\vp_i\}$ is called  analytically stable if
$$\vp_{n+k-1}\circ \vp_{n+k-2} \circ \cdots \circ \vp_n(C)
\not\subset \cali(\vp_{n+k})$$  for any integers $n$ and $k \geq 0$ and any
irreducible
curve on $X_n$, which is equivalent to
$$\vp_{n+k-1}\circ \vp_{n+k-2} \circ \cdots \circ \vp_{n+1}(\calc(\vp_n))
\cap \cali(\vp_{n+k})=\emptyset$$  for any integers $n$ and $k \geq 0$.
\\

The next proposition follows from Prop.~\ref{pr1}.

\begin{prop} \label{pr2} Let $\{\vp_i:X_i\to X_{i+1} \}_{i=0,1,2,\cdots}$  be
analytically stable, then
$$(\vp_{n+k-1}\circ \vp_{n+k-2} \circ \cdots \circ \vp_n)^*(D)  = \vp_n^* \circ
\vp_{n+1}^* \circ \cdots \circ \vp_{n+k-1}^*(D)$$  holds for any divisor $D$ on
$X_{n+k}$.
\end{prop}

\noindent{\it Total transform and proper transform}\\ Let $\pi^{-1}:X \to
Y$ be the
blow-up at the point $p$  and let $D$ be a divisor on $X$. The divisor
$\pi^*(D)$
on $Y$  is called the total transform of $D$. Let $V$ be an analytic
subvariety of
$X$. The closure of the set $\pi^{-1}(V \setminus p)$ in $Y$ is called the
proper
transform of $V$. The divisor $\pi^{-1}(p)$ is also called the total
transform of
the point $p$.\\

\ni{\it Strategy to obtain the space of initial conditions}

Our strategy to obtain the space of initial conditions for a sequence of
rational
mapping is as follows. Let $F(=:Y_{0,i}$, independent of $i)$ be a minimal
surface
and let each
$\vp_i:Y_{0,i}\to Y_{0,i+1}$ be a rational mapping. First blowing up
$Y_{0,i}$ at
the points in
$\cali(\vp_i) \cup \calc(\vp_{i-1})$, we have the surfaces
$Y_{1,i}$ such that each $\vp_i$ can be lifted to a rational holomorphic
mapping
from $Y_{1,i}$ to $Y_{0,i+1}$ and $\vp_{i-1}$ can be lifted to a rational
mapping  from $Y_{0,i-1}$ to $Y_{1,i}$ such that
$\calc(\vp_{i-1})=\emptyset$.  The
mapping $\vp_i$ can also be lifted to a rational mapping from
$Y_{1,i}$ to $Y_{1,i+1}$.  Next by blowing up $Y_{1,i}$ at the points in
$\cali(\vp_i) \cup \calc(\vp_{i-1})$  similarly,  $\vp_i$ is lifted to a
rational
mapping from
$Y_{2,i}$ to $Y_{2,i+1}$.

If we have $Y_{n,i}=Y_{n+1,i}$ for all $i$  for some $n$ by continuing this
operation, then each $\vp_i$ is lifted to a biregular mapping, i.e. an
isomorphism,
from $Y_{n,i}$ to $Y_{n,i+1}$ and hence the sequence of
$X_i:=Y_{n,i}$ can be considered to be the space of initial conditions.  Of
course this
procedure does not terminate for general birational mappings and we may
need not only
blow-ups but also blow-downs for  some mappings.\\

\ni{\it Strategy to obtain a sequence of analytically stable mappings}

Our strategy to obtain a sequence of surfaces such that $\{\vp_i\}$ is
lifted to
an analytically stable system is as follows. Let $F(=:Y_{0,i})$ be a
minimal surface
and let each
$\vp_i:Y_{0,i}\to Y_{0,i+1}$ be a rational mapping.  First blowing up
$Y_{0,i}$ at
the points in
\be \label{ci}
\bigcup_{j=-\infty}^{i-1}
\vp_{i-1}\circ \vp_{i-2} \circ \cdots \circ \vp_{j+1}(\calc(\vp_j))
\cap \cali(\vp_{i}),
\ee  we have the surfaces
$Y_{1,i}$ such that $\{\vp_i\}$ can be lifted to a sequence of  rational
mappings
$\{ \cdots \to Y_{0,i-1}\to Y_{1,i} \to Y_{0,i+1} \}$ such that
$ (\ref{ci})= \emptyset$. Next blowing up $Y_{1,i}$ at the points in
(\ref{ci}) for
$\{\vp_i:Y_{1,i}\to Y_{1,i+1}\}$, we have the surfaces
$Y_{2,i}$ such that $\{\vp_i\}$ can be lifted to a sequence  of rational
mappings
$\{\cdots \to Y_{2,i-1}\to Y_{2,i} \to Y_{1,i+1} \}$ such that
$ (\ref{ci})= \emptyset.$ If we have $Y_{n,i}=Y_{n+1,i}$ for all $i$ for
some $n$ by
continuing this operation, then $\{\vp_i\}$ is lifted to an analytically
stable
sequence of mappings
$\{\vp_i: Y_{n,i}\to Y_{n,i+1}\}$. This procedure does not terminate for
general
sequence of rational mappings but it always terminates for autonomous
systems.\\

\noindent{\it Picard group of rational surface}\\ Let $\pi^{-1}:X \to Y$ be the
blow-up at the point $p$.  Let $C$ be a curve on $X$. The linear
equivalence class
of the total transform of $C$ is written as
$(\pi^{-1})^*[C]$ or simply as $[C]$.  If the point $p$ is an element of
$C$, the
linear equivalence class of the proper transform is $[C]-m[p]$, where $m$
is the
multiplicity of $C$ at $p$.

Let $X$ be a surface obtained by $L$ times blowing up ${\mathbb P}^2$. The
Picard
group $\pic(X)$  is isomorphic to a $\mz $-module in the form
$$\mz E + \sum_{l=1}^L \mz E_l,$$ where $E$ denotes the linear equivalence
class of
the total transform of a generic line in ${\mathbb P}^2$ and $E_l$ denotes
that of
the point of the $l$th blow up.

Let $X$ be a surface obtained by $L$ times blowing up
${\mathbb P}^1\times {\mathbb P}^1$. The Picard group $\pic(X)$,  is
isomorphic to
a $\mz $-module in the form
$$\mz H_0+\mz H_1+ \sum_{l=1}^L \mz E_l,$$ where $H_0$ (or $H_1$) denotes the
linear equivalence class of the total transform of the line $x={\rm constant}$
(resp.
$y={\rm constant}$).\\

\noindent{\it Intersection numbers}\\ Let $X$ be a surface obtained by blow-ups
from ${\mathbb P}^2$. The intersection number of any two divisors on $X$ is
given
by    the following intersection form and their linear combinations,
\begin{eqnarray}\label{isn} E \cdot E = 1,~ E \cdot E_l=0,~ E_l \cdot
E_m=-\delta_{l,m},
\end{eqnarray} where $\delta_{l,m}$ is $1$ if $l=m$ and $0$ if $l\neq m$.

In ${\mathbb P}^1\times {\mathbb P}^1$ case the intersection form is
\begin{eqnarray} H_i \cdot H_j = 1-\delta_{i,j},~ E_l\cdot E_m= -\delta_{l,m},~
H_i \cdot E_l=0~.
\end{eqnarray}

\begin{lemma} \label{le1} Let $X$ and $Y$ be rational surfaces
obtained by blow-ups from $\ptwo$ or $\ponet$. If $\vp: X
\to Y$
is a  rational mapping, the following formulae hold, in the ${\mathbb P}^2$
case
\begin{eqnarray*}
\deg(\vp)=  (\vp)^*(E) \cdot E
\end{eqnarray*} and in the ${\mathbb P}^1\times {\mathbb P}^1$ case
\begin{eqnarray*}
\deg_{x}(P) = (\vp)^*(H_0) \cdot H_1 &
\deg_{y}(P)= (\vp)^*(H_0) \cdot H_0 \\
\deg_{x}(Q)= (\vp)^*(H_1) \cdot H_1 &
\deg_{y}(Q)= (\vp)^*(H_1) \cdot H_0 ,
\end{eqnarray*} where $\vp$ is written as $\vp(x,y)=(P(x,y),Q(x,y))$ in one
of the
local coordinates.\\
\end{lemma}

\bp Let $X'$ be a surface obtained by successive blow-ups
$\pi^{-1}: X \to X'$ that eliminate the indeterminacy of $\vp$ and let
$\vp'$  be the
mapping lifted from $\vp$.  These formulae hold for $\vp'$ (see
\cite{takenawa2}).
The coefficients of $E, H_0$ and $H_1$ on $X$ and on $X'$ are the same and
hence
these formulae also hold for $\vp$.
\ep\\

The next proposition follows from Prop.\ref{pr2} and Lemma,\ref{le1}.

\begin{prop} \label{pr3} Let $X_i$ be rational surfaces obtained by 
blow-ups from $\ptwo$ or $\ponet$ and let
$\{\vp_i:X_i \to X_{i+1}\}$
be a sequence of rational mappings, 
then the following formulae hold,  in the ${\mathbb
P}^2$ case
\begin{eqnarray*}
&&\deg(\vp^n) =(\vp^n)^*(E) \cdot E =\prod_{i=0}^{n-1} \vp_i^*(E) \cdot E\\   
&& \qquad (=\deg(\vp^{-n})= (\vp^n)_*(E) \cdot E,
\mbox{ if $\vp$ is birational})
\end{eqnarray*} where $\vp^n$ denotes
$(\vp_{n-1}\circ \cdots \circ \vp_1\circ \vp_0)$ and in the ${\mathbb
P}^1\times
{\mathbb P}^1$ case
\begin{eqnarray*}
\deg_xP^n(x,y)&=&(\vp^n)^*(H_0)\cdot H_1=(\vp^n)_*(H_1)\cdot H_0\\
\deg_yP^n(x,y)&=&(\vp^n)^*(H_0)\cdot H_0=(\vp^n)_*(H_0)\cdot H_0\\
\deg_xQ^n(x,y)&=&(\vp^n)^*(H_1)\cdot H_1=(\vp^n)_*(H_0)\cdot H_0\\
\deg_yQ^n(x,y)&=&(\vp^n)^*(H_1)\cdot H_0=(\vp^n)_*(H_0)\cdot H_1~.
\end{eqnarray*} where
$P^n(x,y)$ and  $Q^n(x,y)$ denote
$\vp^n(x,y)=(P^n(x,y),Q^n(x,y))$ in one of the local coordinates and the later
equalities hold if $\vp$ is birational.
\end{prop}


We conclude this section by introducing a theorem proposed by  Diller and Favre
\cite{df} which is closely related to our problem.

\begin{prop} Let $X$ be a K\"ahler surface and let $\vp$ be a
birational automorphism of $X$ such that the degree grows linearly in $n$, then
there exists a ruled surface $Y$ such that\\  {\rm i)} $\vp_i$ is birationally
conjugate to a birational automorphism of $Y$.\\ {\rm ii)} there exists a
unique
class $L$ in $\pic(Y)$ which is preserved  by $\vp_i^*$.  $L\cdot L=0$ and
$L$ is a
class of generic fibers which is preserved by $\vp_i$.
\end{prop}

\section{A Projective mapping}

\definition (confining singularity pattern) Let $X_i$'s be smooth projective
surfaces and
$\{\vp_i:X_i\to X_{i+1}\}$ be a sequence of rational mappings. Some effective
divisor topologically may appear or disappear for some of these mappings and
we call these sequences singularity patterns. Singularity patterns can be
classified
as follows,  (EF means effective divisor) \\ i) $\cdots$$\to$ EF $\to$
EF$\to$ point
$\to$
$\cdots$ $\to$ point$\to$ EF
$\to$ EF $\to$ $\cdots$: {\it strictly confining},\\ ii) $\cdots$ $\to$
point$\to$
point$\to$ EF $\to$ $\cdots$ $\to$ EF$\to$ point
$\to$ point $\to$ $\cdots$: {\it confining} (but not strictly confining),\\
iii)
othrewise: {\it non-confining}.\\

In this section we consider a mapping which is confining but not strictly
confining and   whose degree is constant with respect to the iteration. We
shall
illustrate how to use the method of space of initial conditions in this
case. It turns
out that the corresponding space of initial conditions needs an infinite
number of
blow-ups, or sometimes blow-downs, but still has a meaning.

Let us consider the following simple non-autonomous discrete dynamical system:
\be x_{n+1}=-a_n-\frac{b_n}{x_n}-\frac{1}{x_n x_{n-1}}
\ee where $a_n$ and $b_n$ are free functions of $n\in \mz $. This equation is
equivalent to the following 2-dimensional mapping
$\vp_n:(x,y)\in {\mathbb C}^2 \mapsto  (\ol{x},\ol{y})\in{\mathbb C}^2$:
\be \vp_n:
\left\{ \ba{lll}
\ol{x}&=&y\\
\ol{y}&=&(-a_n x y-b_n x-1)/xy
\ea \right. \label{proj11}
\ee where $(\overline{x},\overline{y})$ means the image of $(x,y)$ by $\vp_n$.

Let us consider $\vp_n$ to be a mapping from  the complex projective space
$\ponet$ to itself. We take a coordinate system of ${\mathbb P}^1 \times
{\mathbb
P}^1$ as
$U_1 \ni (x,y), U_2 \ni (x,Y), U_3 \ni(X,y), U_4\ni(X,Y)$ where $X=1/x,
Y=1/y$ and
$U_i\cong {\mathbb C}^2$. In these coordinates $\vp_n$ is written as
follows (for
simplicity we write $\ol{y}$ only)
\be \label{proj2}
\ol{y}=\left\{ \ba{ll} (-a_nxy-b_nx-1)/xy & \mbox{ from }U_1\\
(-a_nx-b_nxY-Y)/x &
\mbox{ from }U_2\\ (-a_ny-b_n-X)/y & \mbox{ from }U_3\\ -a_n-b_nY-XY &
\mbox{ from
}U_4
\ea \right.
\ee From Eq.(\ref{proj2}) we can find the indeterminate points of $\vp_n$ (the
points where $\ol{y} \in {\mathbb P}^1$ is indeterminate):
$(x,y)=(-1/b_n,0)$ and $(0,\infty)$.

\subsection{Space of initial conditions (by blow-ups)}\label{pro1}

\noindent{\it Blow-up of a rational surface}\\ We denote the blow-up at
$(x,y)=(x_0,y_0)\in {\mathbb C}^2$:
\begin{eqnarray*} &&\{(x,y):x,y \in {\mathbb C}\}\\
&\mapleft{\mu_{(x_0,y_0)}} &
\{(x-x_0,~y-y_0;~\zeta_1:\zeta_2)~|~x,y,\zeta_1,\zeta_2 \in  {\mathbb C},\\
&&\quad
|\zeta_1|+|\zeta_2|\neq 0, (x-x_0)\zeta_2=(y-y_0)\zeta_1 \}
\end{eqnarray*} by
\begin{eqnarray}\label{coblow}
 (x,y) \leftarrow (u,v)=(x-x_0,\frac{y-y_0}{x-x_0})\cup
(u',v')=(\frac{x-x_0}{y-y_0}, y-y_0).
\end{eqnarray}

First we blow up at $(x,Y)= (0,0)$,
$(x,Y) \leftarrow (x,Y/x)\cup (x/Y,Y)$ and denote the obtained surface by
$Y_n'$.
Then
$\vp_n$ is lifted to a birational mapping  from $Y_n'$ to ${\mathbb P}^1 \times
{\mathbb P}^1$. For example, in the new coordinates $\vp_n$ is expressed as
\ben  (u_E,v_E):=(x,Y/x) &\mapsto& (\overline{x},\overline{y})= (1/u_E v_E,
-a_n-b_nu_E v_E-v_E)\\ (u_E',v_E'):=(x/Y,Y) &\mapsto&
(\overline{x},\overline{y})=(1/v_E',
    (-a_nu_E'-b_nu_E'v_E'-1)/u_E')
\een (use (\ref{proj11}) and the definition of $u_E, v_E, u_E'$ and
$v_E'$). The
exceptional curve is described as $u_E=0$ (or $v_E'=0$) and its image is
described
as
$(\ol{X},\ol{y})=(0, -a_n-v_E)$ ($\ol{X}=1/\ol{x}$). 
In this way, blowing up at
$(x,y)=(x_0,y_0)$
gives meaning  to $(x-x_0)/(y-y_0)$ at this point.\\

The indeterminate points of the inverse mapping $\vp_{n-1}^{-1}$:
\be \vp_{n-1}^{-1}:
\left\{ \ba{lll}
\ul{x}&=&-1/(x y+a_{n-1} x+b_{n-1})\\
\ul{y}&=&x
\ea \right.,
\ee are $(x,y)=(\infty,-a_{n-1})$ and $(0,\infty)$, where $(\ul{x},\ul{y})$
means the
image of $(x,y)$ by
$\vp_{n-1}^{-1}$. By successive blow-ups we can eliminate the indeterminacy of
$\vp_n$ and $\vp_{n-1}^{-1}$ and we obtain  the surface $Y_{1,n}$ defined
by the
following sequence  of blow-ups (for simplicity  we write only one of the
coordinates of (\ref{coblow})):
\ben (x,y) &\maplleft{\mu_E}{(0,\infty)}& (u_E,v_E)=
\left(x,\frac{1}{xy}\right)\\ (x,y) &\maplleft{\mu_{F_1}}{(-\frac{1}{b_n},0)}&
(u_{F_1},v_{F_1})=
\left(x+\frac{1}{b_n},\frac{y}{x+1/b_n}\right) \\ (x,y)
&\maplleft{\mu_{G_1}}{(\infty,-a_{n-1})}&  (u_{G_1},v_{G_1})=
\left(\frac{1}{x},x(y+a_{n-1})\right).
\een and $\vp_n$ (and $\vp_{n-1}^{-1}$) is lifted to a regular mapping from
$Y_{1,n}$  to ${\mathbb P}^1\times{\mathbb P}^1$.

\begin{figure}[ht]
\begin{center}
\includegraphics*[width=8cm,height=5cm]{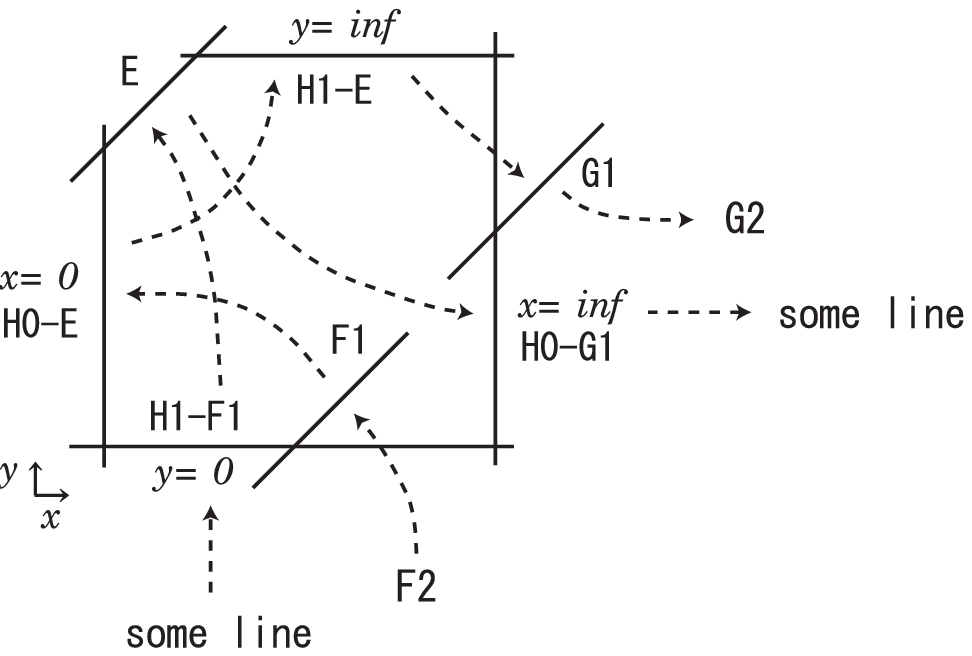}
\caption[]{Sect.\ref{pro1}}\label{projective}
\end{center}
\end{figure}

Following the scheme of the previous section, we successively blow up at
the points
each of which is the image or the pre-image of some curve. We assume  the
points
$\vp_{k-i}^{-1}\circ \vp_{k-2}^{-1}\circ \cdots\circ  
\vp_{k-1}^{-1} (-\frac{1}{b_{k}},0)$ or
$\vp_{k+i}\circ \vp_{k+2}\circ \cdots \circ \vp_{k+1}(\infty,-a_k)$  do not
meet the lines $x=0,x=\infty,y=0,y=\infty$ for any integers $k$ and $i\leq 1$. For generic $\{a_n\}$ and $\{b_n\}$
this is true.  Blowing-up at these points, we obtain the sequence of ``rational
surfaces''  $\{X_n:=Y_{\infty,n}\}$.  (See Fig.\ref{projective}).

We denote the class of the total transform of the line $x={\rm constant}$ or
$y={\rm constant}$ as $H_0$ and $H_1$ respectively and denote the class of the
total transform of the points
$(0,\infty),
\vp_{k-i}^{-1}\circ  \vp_{k-2}^{-1}\circ \cdots \circ 
\vp_{k-1}^{-1} (-\frac{1}{b_{k}},0)$ or
$\vp_{k+i}\circ \vp_{k+2}\circ \cdots \circ \vp_{k+1}(\infty,-a_k)$ 
($i=0,1,2,\cdots$) as $E,F_i$
or $G_i$ respectively. In Fig.\ref{projective}  each line  means an irreducible
curve and the intersection of lines means the intersection of corresponding
curves.
For example the class of total transform of the line $x=0$ is $H_0$ and the
class
of its proper transform is  $H_0-E$ (see ``Picard group of  rational
surfaces'' in
Section \ref{pre}).\\

\remark Of course the surface $X_n$ is defined by infinite times blowing up and
therefore is not a rational surface, but we can justify our discussion
through the
arguments below. We denote the blow-ups of
$E,F_i$ and  $G_i$ as $\mu_{E},\mu_{F_i}$ and  $\mu_{G_i}$ respectively.
Let $N$
be a sufficiently large positive integer and let $k$ be an integer such that
$|k|<N$.  Let $X_{k,n}$ be a rational surface obtained by  blow-ups
$\mu_{E},\mu_{F_1},\mu_{F_2},\cdots,\mu_{F_{N-k}}$ and
$\mu_{G_1},\mu_{F_2},\cdots,\mu_{G_{N+k}}$. Then $\vp_n$ is lifted to an
isomorphism from $X_{k,n}$ to $X_{k-1,n+1}$.\\

\ni{\it Actions on the Picard group}

The Picard group of $X_n$ is isomorphic to the lattice
$$\mz H_0 + \mz H_1+\mz E+\sum_{i=1}^{\infty}\mz F_i
  +\sum_{i=1}^{\infty}\mz G_i.$$ The mapping $\vp_n$ maps $\pic(X_n)$ to
$\pic(X_{n+1})$ as
\ben H_1&\mapsto&H_0 ~~(\mbox{if $y$ is a constant, then $\ol{x}$ is a
constant})\\
H_1-E&\mapsto&G_1\\ H_0-E&\mapsto&H_1-E\\ E&\mapsto&H_0-G_1\\
F_1&\mapsto&H_0-E\\
H_1-F_1&\mapsto&E\\ F_i&\mapsto&F_{i-1} ~~(\mbox{for } i\geq 2)\\
G_i&\mapsto&G_{i+1}
\een and hence the linear transformation $(\vp_n)_*:\pic(X_n)\to
\pic(X_{n+1})$ is
\ben H_0&\mapsto&H_0+H_1-E-G_1\\ H_1&\mapsto&H_0\\  E&\mapsto&H_0-G_1\\
F_1&\mapsto&H_0-E\\ F_i&\mapsto&F_{i-1} ~~(\mbox{for } i\geq 2)\\
G_i&\mapsto&G_{i+1}.
\een

By using Prop. \ref{pr3} and
$$\left(\prod_{k=0}^{m-1} (\vp_{n+k})\right)_*(H_0)= H_0+H_1-E-G_m$$
$$(\vp_{n+k})_*(H_1) = H_0,$$ the degree of $(\prod_{k=0}^{m-1} \vp_{n+k})(x,y)
=(P^m,Q^m) $, $(m\geq 1)$, can be calculated as
\ben &&\deg_x P_m=1 ~(\mbox{if } m\geq 2),~ 0 ~(\mbox{if } m=1), \\ &&\deg_y
P_m=\deg_x Q_m=\deg_y Q_m=1.
\een

\subsection{Space of initial conditions (by blow-ups and
blow-downs)}\label{pro2}

\begin{figure}[ht]
\begin{center}
\includegraphics*[width=8cm,height=5cm]{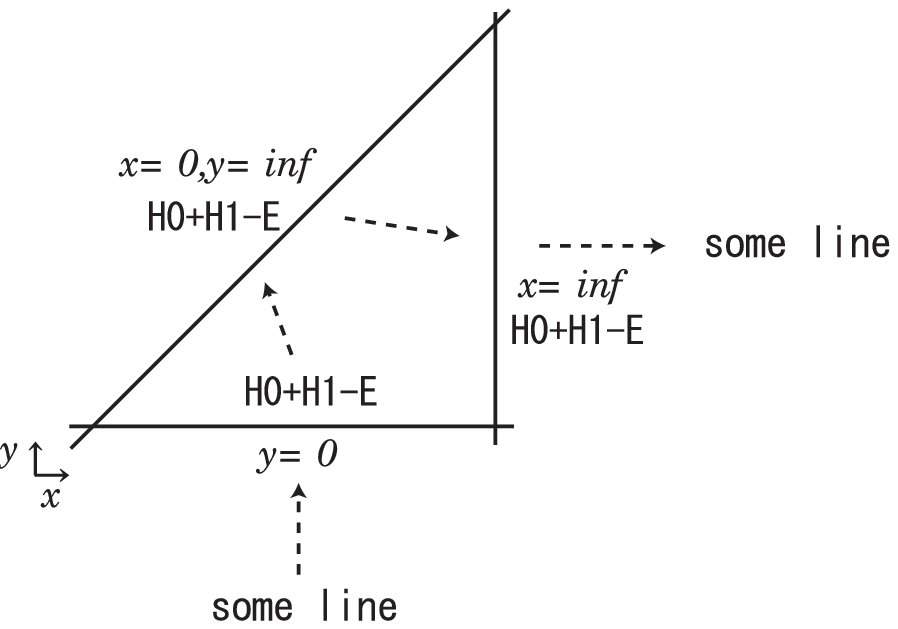}
\caption[]{Sect.\ref{pro2}}\label{projective2}
\end{center}
\end{figure}

From Fig.\ref{projective} we can find the fact that
$\vp_n$ is lifted to an automorphism on the surface obtained by the blow-up
$\mu_E$
and blow-downs along the proper transforms  of the lines $x=0$ and $y=0$. This
surface is nothing but $\ptwo$ (see Fig.\ref{projective2}).

The concrete calculation is as follows.
\ben &&(x,y)\cup(x,\frac{1}{y})\cup(\frac{1}{x},y)\cup
(\frac{1}{x},\frac{1}{y})
=\ponet\\ &\leftarrow&
(x,y)\cup(x,\frac{1}{xy})\cup(xy,\frac{1}{y})\cup(\frac{1}{x},y)\cup
(\frac{1}{x},\frac{1}{y})\\ &\rightarrow&
(x,xy)\cup(x,\frac{1}{xy})\cup(\frac{1}{x},y)\cup(\frac{1}{x},\frac{1}{y})\\
&\rightarrow& (x,xy)\cup(\frac{1}{x},y)\cup(\frac{1}{y},\frac{1}{xy})\\
&=&(x,z)\cup(\frac{1}{x},y)\cup(\frac{1}{y},\frac{1}{z}) =\ptwo ~~(z=xy).
\een By putting $x=v/u,y=w/v,z=w/u$, $\vp_n$ reduces to
the mapping from $\ptwo$ to itself:
\be (\ol{u}:\ol{v}:\ol{w})&=&(v:w:-a_nw-b_nv-u)
\ee and therefore  $\vp_n$ has been reduced to a linear  transformation on
$\ptwo$.\\

\remark  The singularity pattern of $\{\vp_n:\ponet \to \ponet\}$
corresponding to
$F_i$ and $G_i$ is
$\{ \cdots$, point, point, line, line, point, point, $\cdots\}$.  In these case
there is a possibility that the lines can be blown down (but not always, see
Section 3 and 4).

\subsection{Analytically stable surface}\label{pro3}

\begin{figure}[ht]
\begin{center}
\includegraphics*[width=8cm,height=5cm]{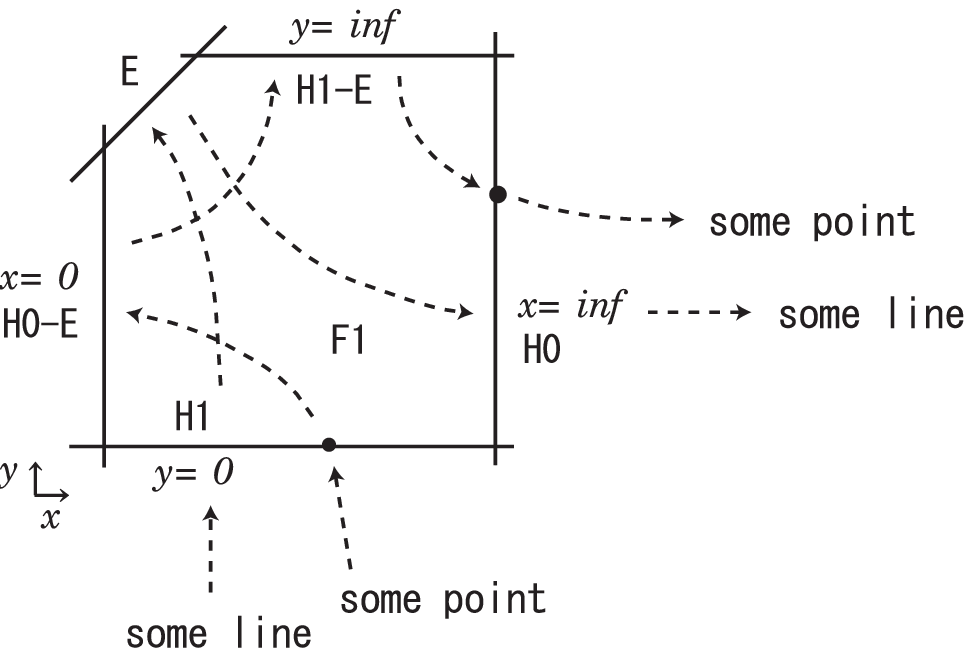}
\caption[]{Sect.\ref{pro3}}\label{projective3}
\end{center}
\end{figure}

Following the scheme in Section~2 we can obtain a surface such that
$\{\vp_n\}$ is
analytically stable. Recall that a sequence of mappings $\{\psi_i:X_i\to
X_{i+1}\}$
is analytically stable if and only if there does not exist a singularity
pattern
[EF $\to$ point $\to \cdots \to$ point
$\to$ EF], where EF means effective divisor, and hence we can obtain such a
sequence surfaces by blowing up at all the points in these singularity
patterns of
$\{\vp_n:\ponet \to \ponet \}$.   Let $X$ be the surface obtained by the
blow-up
$\mu_E$, then $\{\vp_n:X\to X\}$ is analytically stable.  In fact, since
$\cali(\vp_n)=\{(-1/b_n,0)\})$ and
${\cal C}(\vp_m)=\{(\infty,-b_m)\}$ we have
$\cali(\vp_n) \cap (\prod_{k=1}^{n-m-1}\vp_{m+k})({\cal
C}(\vp_m))=\emptyset$ (we have
assumed that $a_n$ and $b_n$ are generic)(see Fig.\ref{projective3}).

The action $(\vp_n)_*$ on $\pic(X)$ is
\ben H_0&\mapsto&H_0+H_1-E\\ H_1&\mapsto&H_0\\  E&\mapsto&H_0.
\een


\section{A simple Riccati system}

In this section and the next one we consider some mappings which are
non-confining
but  whose degrees grow linearly with the iteration of these mappings.
These systems
can be linearised to ``Riccati'' systems as explained in \cite{rog}. We
shall show
that this linearisation can be recovered in a straightforward way  by using
the method
of the space of initial conditions.

In this section let us consider the following simple non-autonomous discrete
dynamical system:
\be \label{ric} x_{n+1}&=&(-\frac{x_n}{x_{n-1}}+a_n)x_n
\ee This mapping is linearised in a straightforward way, since it can be
written as
$$w_n=-w_{n-1}+a_n, \qquad  x_{n+1}=x_n w_n$$ by putting $w_n=x_{n+1}/x_n$. This
system can be solved in cascade (the former equation is linear (projective in
general) and the later  one is linear (resp. projective) with respect to
$x_n$).
We will recover this transformation by using the space of initial conditions.

Eq.~(\ref{ric}) reduces to the mapping $\vp_n:\ponet \to \ponet$:
\be \label{ric1}
\left\{ \ba{lll}
\ol{x}&=&y\\
\ol{y}&=&\ds (-\frac{y}{x}+a_n)y
\ea \right.
\ee and the indeterminate points of $\vp_n$ and $\vp_{n-1}^{-1}$ are
$(x,y)=(0,0)$ and $(\infty,\infty)$.

\subsection{Space of initial conditions}\label{ricc1}

\begin{figure}[ht]
\begin{center}
\includegraphics*[width=8cm,height=5.5cm]{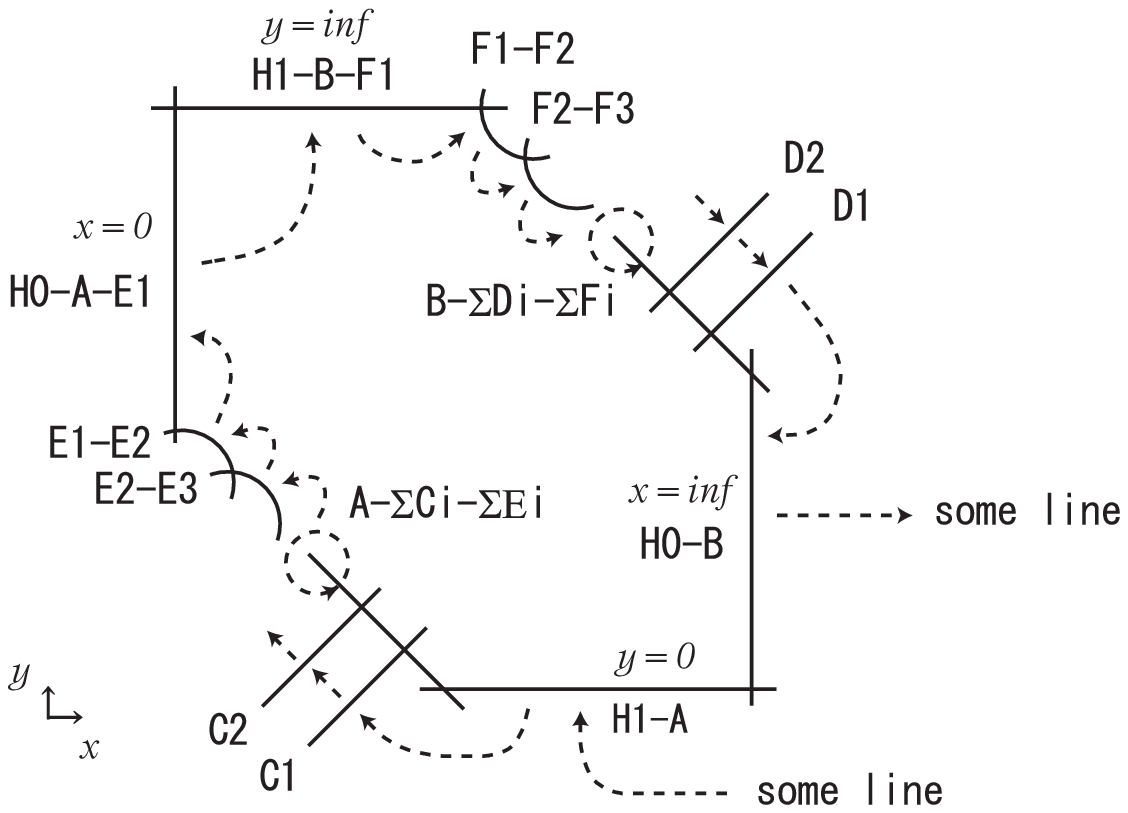}
\caption[]{Sect.\ref{ricc1}}\label{riccati}
\end{center}
\end{figure}

In the same way as in Section~\ref{pro1} we obtain the sequence of  ``rational
surfaces'' $X_n$ as Fig.\ref{riccati}.

In the surface $X_n$ the total transforms of the points of blow-ups are  as
follows:
\ben &&A: (x,y)=(0, 0)\\ &&B: (1/x,1/y):=(0,0)\\ &&C_i: (u_A,v_A):=(x,y/x)=(0,~
\sum_{k=1}^i a_{n-i-1})\\ &&D_i: (u_B,v_B):=(y/x,1/y)=(-\sum_{k=0}^{i-1}
a_{n+i},~
0)\\ &&E_1: (u_A',v_A'):=(x/y,y)=(0,0)\\ &&E_2:
(u_{E_1},v_{E_1}):=(u_A',v_A'/u_A')=(0,0)\\ &&E_{i+1}:
(u_{E_i},v_{E_i}):=(u_{E_{i-1}},v_{E_{i-1}}/u_{E_{i-1}})=(0,0)
\quad (i \geq 2)\\ &&F_1: (u_B',v_B'):=(1/x,x/y)=(0,0)\\ &&F_2:
(u_{F_1},v_{F_1}):=(u_B'/v_B',v_B')=(0,0)\\ &&F_{i+1}:
(u_{E_i},v_{E_i}):=(u_{E_{i-1}}/v_{E_{i-1}},v_{E_{i-1}})=(0,0)
\quad (i \geq 2).
\een

The action $(\vp_n)_*:\pic(X_n)\to \pic(X_{n+1})$ is
\ben H_0&\mapsto&2H_0+H_1-A-B-C_1-F_1\\ H_1&\mapsto&H_0\\  A&\mapsto&H_0-C_1\\
B&\mapsto&H_0-F_1\\ C_i&\mapsto&C_{i+1} \quad (i \geq 1)\\ D_1&\mapsto&H_0-B\\
D_i&\mapsto&D_{i-1} \quad (i \geq 2)\\ E_1&\mapsto&H_0-A\\ E_i&\mapsto&E_{i-1}
\quad (i \geq 2)\\ F_i&\mapsto&F_{i+1} \quad (i \geq 1).
\een

The degree of $(\prod_{k=0}^{m-1} \vp_{n+k})(x,y)=(P^m,Q^m)$, $(m\geq 1)$,
can be
calculated by
$$\prod_{k=0}^{m-1} (\vp_{n+k})_*(H_0)=
 (m+1)H_0+m(H_1+A+B)-\sum_{i=1}^m C_i-\sum_{i=1}^m F_i$$
$$(\vp_{n+k})_* (H_1)=H_0$$ as
$\deg_xP^m=m-1,~ \deg_yP^m=m,~ \deg_xQ^m=m,~ \deg_yQ^m=m+1$.

\subsection{Analytically stable surface}\label{ricc2}

\begin{figure}[ht]
\begin{center}
\includegraphics*[width=8cm,height=5cm]{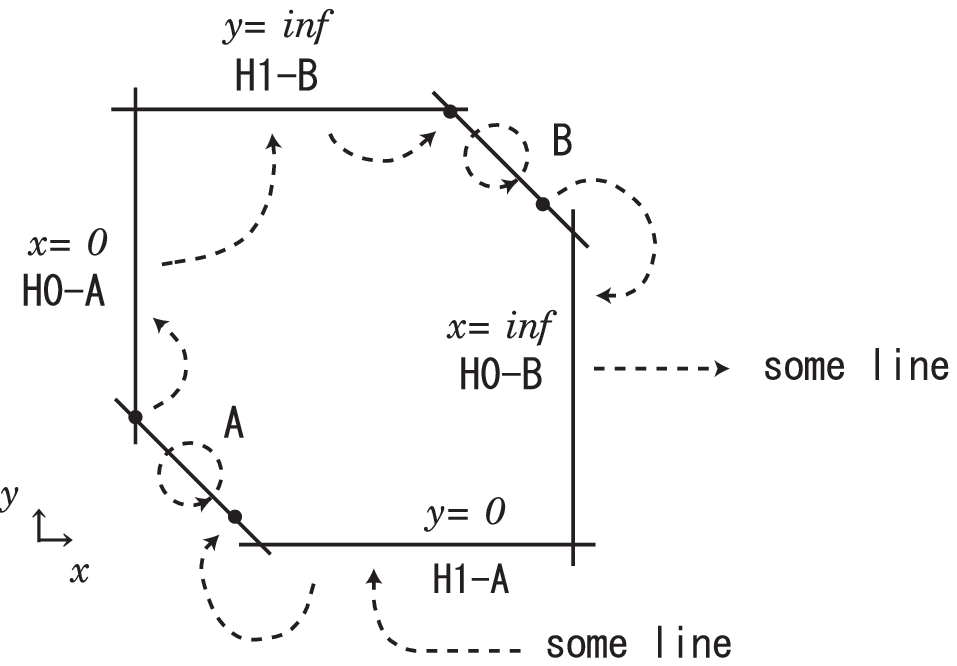}
\caption[]{Sect.\ref{ricc2}}\label{riccati2}
\end{center}
\end{figure}

Similar to Section\ref{pro3}, the sequence of mappings $\{\vp_n\}$ is
lifted to an
analytically stable sequence. Let $X$ be the surface obtained by the blow-ups
$\mu_A$ and $\mu_B$, then $\{\vp_n:X\to X\}$ is analytically stable (see
Fig.\ref{riccati2}). Notice that $X$ is independent of $n$, while this is
not true for the more complicated Riccati mappings as in Sect.5.

The action $(\vp_n)_*:\pic(X)\to \pic(X)$ is
\ben H_0&\mapsto&2H_0+H_1-A-B\\ H_1&\mapsto&H_0\\  A&\mapsto&H_0\\
B&\mapsto&H_0.
\een

The invariant effective class in $\pic(X)$ is
\be \label{ricinv} H_0 + H_1-A-B
\ee and the divisors in this class, i.e. divisors in $\ponet$ such that the
class
of their proper transforms is (\ref{ricinv}),  are
$$c_1 x+c_2 y=0,$$ where $c_1$ and $c_2$ are nonzero constants in $\mc$.
Actually
these lines pass through the points  $(x,y) = (0,0),(\infty,\infty)$ with
multiplicity
$1$. Here $\{(c_1:c_2)\} \simeq \pone$ and the  lines $c_1 x+c_2 y=0$ can be
considered as the base space and as fibers respectively.  Hence $X$ has a
fibration
preserved by
$\vp_n$.

By putting $u=-c_1/c_2=y/x, v=x$ $\vp$ is decomposed to mappings on  the
base space
and on fibers as follows
\be \left\{ \ba{lll}
\ol{u}&=&u+a_n\\
\ol{v}&=&uv .
\ea \right.
\ee


\section{ A more complicated Riccati system}

In this section we consider the mapping
\begin{eqnarray}\label{rog}
1=\frac{f_{n-1}+f_{n+1}+k}{x_n}-\frac{f_{n-1}+f_n+k}{x_n+x_{n-1}}
 -\frac{f_n+f_{n+1}+k}{x_n+x_{n+1}},
\end{eqnarray} where $f_n$'s are free functions of $n$ and $k$ is a constant in
$\mc$ (hence we can assume $k=0$ without loss of generality). This mapping is
obtained by limiting process from the discrete Painlev\'e  IV equation
\cite{rog} and
the degree grows linearly. Eq.~(\ref{rog}) reduces to the mapping
$\vp_n:\ponet \to
\ponet$:
\be \label{rog1}
\left\{ \ba{lll}
\ol{x}&=&y\\
\ol{y}&=& \ds - \frac{y(xy+y^2-(f_{n-1}-f_n)x+2f_n y)}
{xy+y^2-(f_{n-1}+f_{n+1})x+(f_n-f_{n+1})y}
\ea \right.
\ee and the indeterminate points of $\vp_n$ and $\vp_{n-1}^{-1}$ are
$(x,y)=(0,0)$ and $(\infty,\infty)$.

\subsection{Space of initial conditions}\label{crs1} 
\begin{figure}[ht]
\begin{center}
\includegraphics*[width=8cm,height=5.5cm]{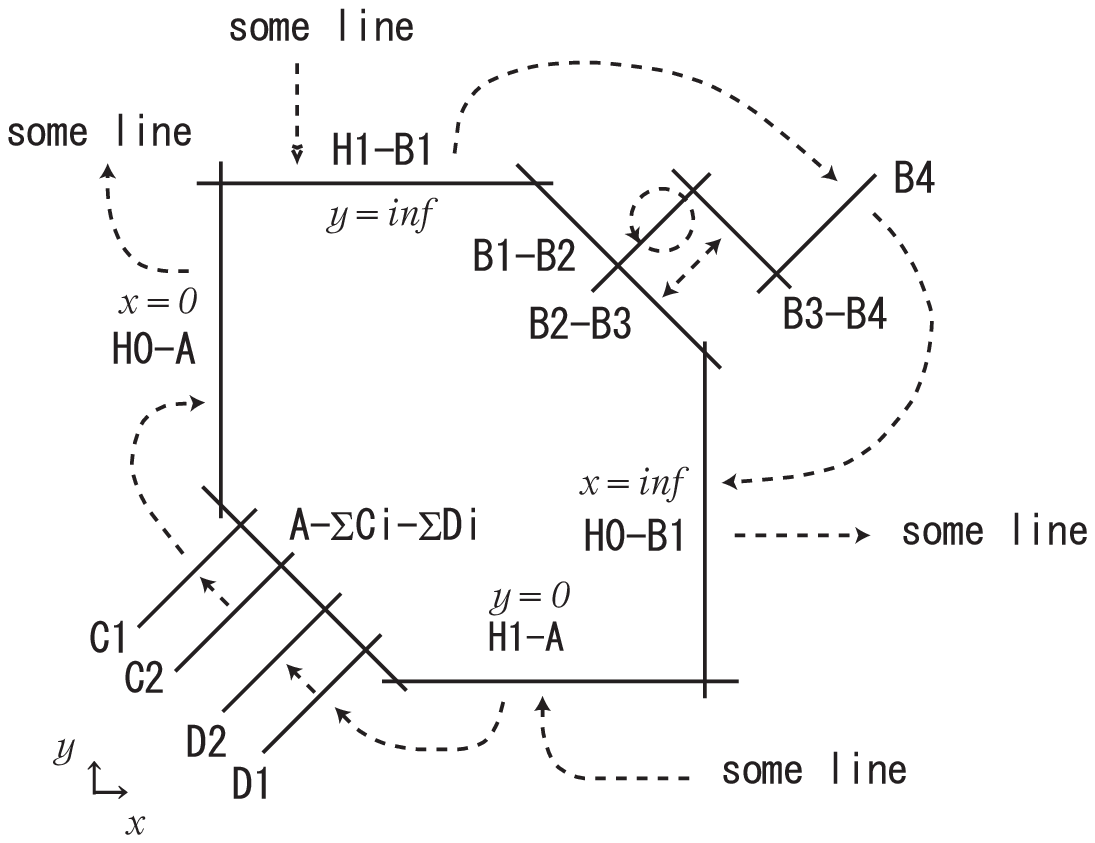}
\caption[]{Sect.\ref{crs1}}\label{new}
\end{center}
\end{figure}

``The rational
surfaces'' $X_n$
corresponding to $\vp$ for generic $\{f_n\}$  is as Fig.\ref{new}. \\

In the surface $X_n$ the total transforms of the points of blow-ups are  as
follows
\ben &&A: (x,y)=(0, 0)\\ &&B_1: (1/x,1/y):=(0,0)\\ &&B_2:
(u_{B_1},v_{B_1}):=(1/x,x/y)=(0,-1)\\ &&B_3:
(u_{B_2},v_{B_2}):=(u_{B_1},\frac{v_{B_1}+1}{u_{B_1}}) =(0,f_{n-1}+f_n))\\
&&B_4:
(u_{B_3},v_{B_3}):=(u_{B_2},\frac{v_{B_2}-(f_{n-1}+f_n)}{u_{B_2}})
=(0,-(f_{n-1}+f_n)2f_{n-1})\\ &&C_1:
(u_A,v_A):=(x,y/x)=(0,\frac{f_{n-1}+f_{n+1}}{f_n-f_{n+1}}=:c_{1,n})\\ &&C_i:
(u_A,v_A)= (0,\frac{(f_{n-1}-f_n)+(f_{n-1}+f_{n+1})c_{i-1,n+1}}{
2f_n+(f_n-f_{n+1})c_{i-1,n+1}}=:c_{i,n}) \quad(i\geq 2)\\ &&D_1: (u_A,v_A)=(0,
\frac{f_{n-1}-f_{n-2}}{f_n+f_{n-2}}=:d_{1,n})\\ &&D_i: (u_A,v_A)=
(0,\frac{(f_{n-1}-f_{n-2})+(2f_{n-1})d_{i-1,n-1}}{
 (f_{n-2}+f_n)+(f_n-f_{n-1})d_{i-1,n-1}}=:d_{i,n})
                              \quad(i\geq 2)\\
\een

The action $(\vp_n)_*:\pic(X_n)\to \pic(X_{n+1})$ is
\be H_0&\mapsto&3H_0+H_1-A-B_1-B_2-B_3-B_4-D_1 \nn \\ H_1&\mapsto&H_0\nn\\
A&\mapsto&H_0-D_1\nn\\ B_1&\mapsto&H_0-B_4\nn\\ B_2&\mapsto&H_0-B_3\nn\\
B_3&\mapsto&H_0-B_2 \label{goract} \\ B_4&\mapsto&H_0-B_1\nn\\
C_1&\mapsto&H_0-A\nn\\ C_i&\mapsto&C_{i-1} \quad (i \geq 2)\nn\\
D_i&\mapsto&D_{i+1} \quad (i \geq 1).\nn
\ee The degree of $(\prod_{k=0}^{m-1} \vp_{n+k})(x,y)=(P^m,Q^m)$, $(m\geq
1)$,  can
be calculated by
$$\prod_{k=0}^{m-1} (\vp_{n+k})_*(H_0)=  (2m+1)H_0+(2m-1)(H_1-A)- m
\sum_{l=1}^4B_l
-2\sum_{l=1}^{m-1}D_l -D_m$$
$$(\vp_{n+k})_*(H_1)=H_0$$  as
$$\deg_xP=0,~ \deg_xP^m=2m-3 (m\geq 2),$$
$$\deg_y P^m=2m-1,~ \deg_xQ^m=2m-1,~ \deg_yQ^m=2m+1 (m\geq 1).$$

\subsection{Analytically stable surfaces}\label{crs2}

\begin{figure}[ht]
\begin{center}
\includegraphics*[width=8cm,height=5.5cm]{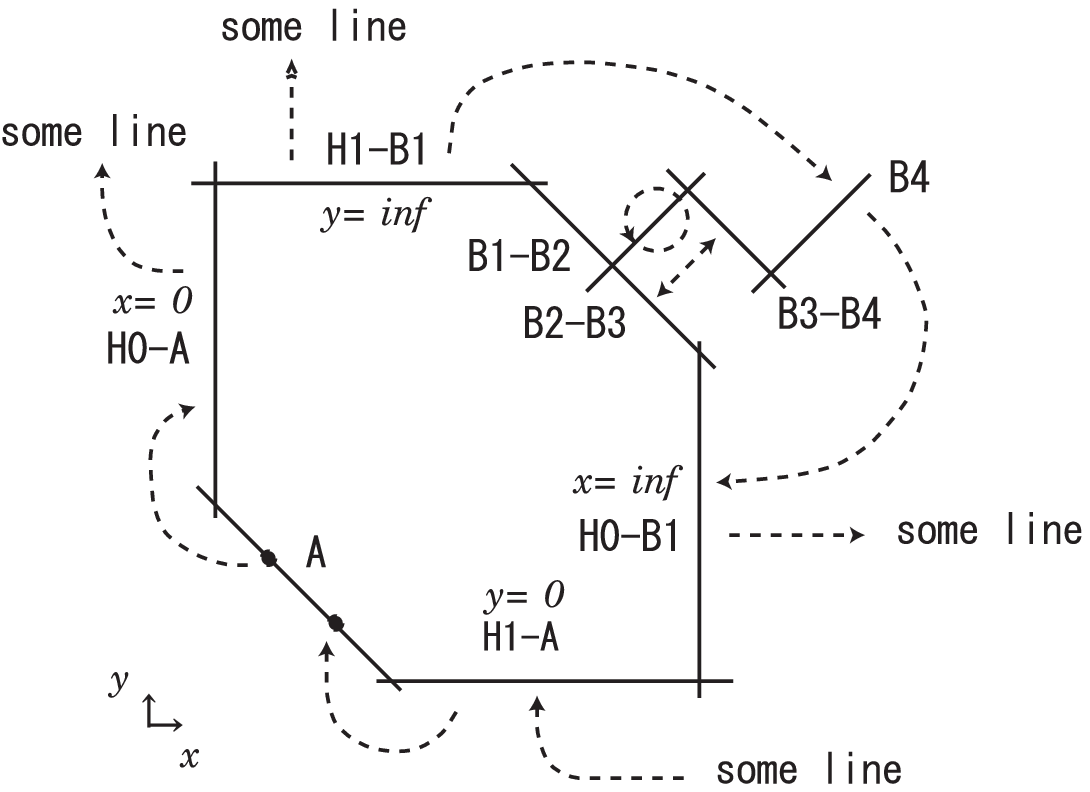}
\caption[]{Sect.\ref{crs2}}\label{new2}
\end{center}
\end{figure}

Let $X_n$ be the surface obtained by the blow-ups
$\mu_A$ and $\mu_{B_l} (l=1,2,3,4)$, then
$\{\vp_n:X_n\to X_{n+1}\}$ is analytically stable (see Fig.\ref{new2}).

The action $(\vp_n)_*:\pic(X_n)\to \pic(X_{n+1})$ is
\ben H_0&\mapsto&3H_0+H_1-A-B_1-B_2-B_3-B_4\\ H_1&\mapsto&H_0\\
A&\mapsto&H_0\\
B_1&\mapsto&H_0-B_4\\ B_2&\mapsto&H_0-B_3\\ B_3&\mapsto&H_0-B_2\\
B_4&\mapsto&H_0-B_1.
\een

The invariant effective class in $\pic(X_n)\simeq \pic(X_{n+1})$ is
$$2H_0 + 2H_1-2A-B_1-B_2-B_3-B_4$$  and the divisors in this class are
\be \label{curve} c_1 (x+y)^2 + c_2 y(x^2-f_n y+xy+f_{n-1}(2x+y))=0
\ee where $c_1$ and $c_2$ are nonzero constants in $\mc$. Here
$\{(c_1:c_2)\} \simeq
\pone$ and the  curves (\ref{curve}) can be considered as the base space
and as fibers
respectively.  Hence $X$ has a fibration preserved by $\vp_n$. Here the curve
(\ref{curve}) has singular point $(0,0)$ and is birationally isomorphic to
$\pone$. Actually the line $sx-y=0$ intersects with the curve (\ref{curve})
at four
points in $\ponet$: $(0,0)$ (order 2), $(\infty,\infty)$ and
\be \label{curve2} \left\{ \ba{lll} x&=& \ds -\frac{c(1+s)^2+s(-f_n
s+f_{n-1}(2+s))}{s(1+s)}\\  y&=& \ds -\frac{c(1+s)^2+s(-f_n
s+f_{n-1}(2+s))}{(1+s)},
\ea \right. \ee where $c=c_1/c_2$. Eq.~(\ref{curve2}) gives a birational
isomorphism from $s \in \pone$  to the curves (\ref{curve}).

Using the new independent variables $c$ and $s$, $\vp$ reduces to
\be \left\{ \ba{lll}
\ol{c}&=& \ds \frac{f_{n-1}+f_n}{f_n+f_{n+1}}(c+f_{n-1}-f_n) \\
\ol{s}&=& \ds -\frac{(c+f_{n-1}-f_n)(s+1)}{(c+f_{n-1}-f_n)s+c+f_{n-1}+f_{n+1}}.
\ea \right. \ee  Since $d:=(f_{n-1}+f_n)c+f_{n-1}^2$ is a constant,  the first
equation is easily integrated and the second equation is a projective
mapping on
$\pone$.\\

\subsection{Some autonomous case}\label{crs3} 
\begin{figure}[ht]
\begin{center}
\includegraphics*[width=8cm,height=5.5cm]{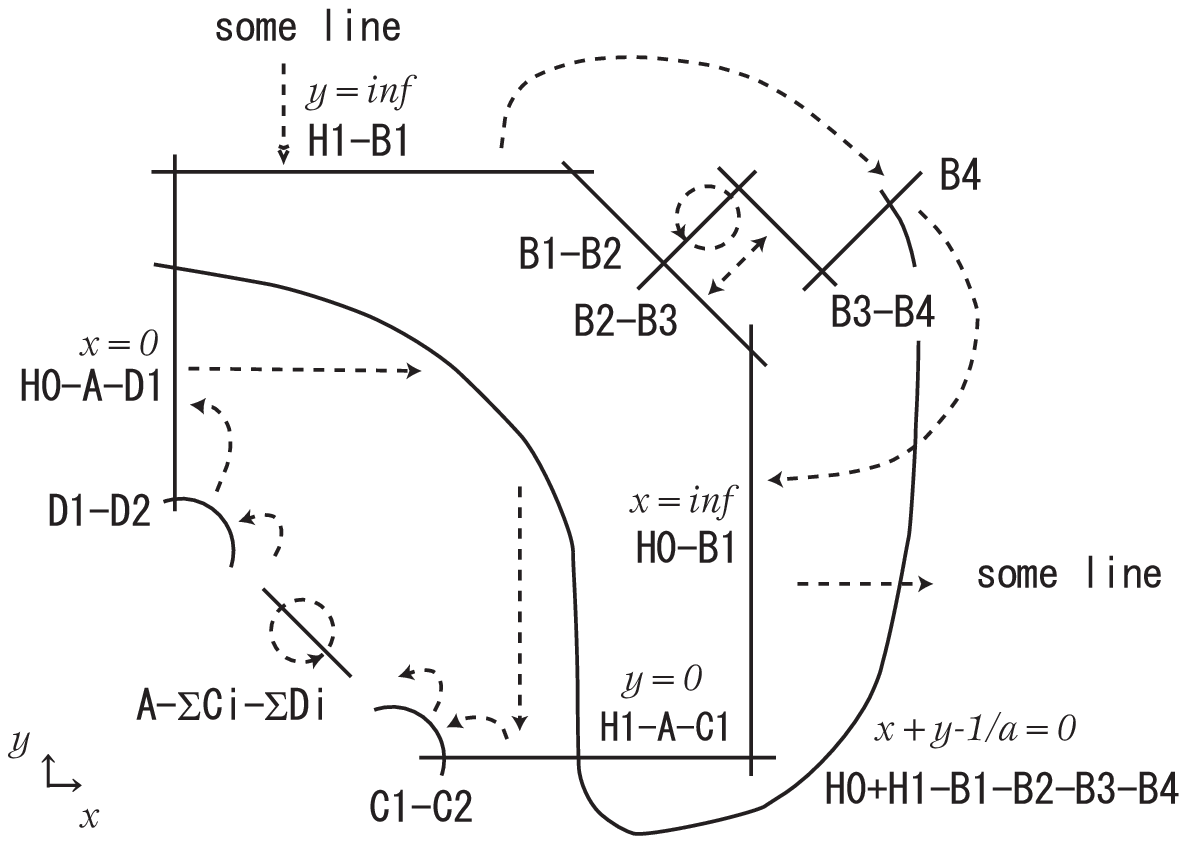}
\caption[]{Sect.\ref{crs3}}\label{new3}
\end{center}
\end{figure}

In some autonomous cases the
space of
initial conditions is different from the generic case. We consider the case
where
$f_n=-1/2a$ for all $n$ in Eq.~(\ref{rog1}):
\be \label{rog3}
\left\{ \ba{lll}
\ol{x}&=&y\\
\ol{y}&=& \ds - \frac{y^2(a x+ a y - 1)} {a xy+a y^2 + x}
\ea \right.
\ee ``The rational surfaces'' $X_n$ corresponding to Eq.(\ref{rog3})  is as
Fig.\ref{new3}.

In the surface $X_n$ the total transforms of the points of blow-ups are  as
follows
\ben &&A: (x,y)=(0, 0)\\ &&B_1: (1/x,1/y):=(0,0)\\ &&B_2:
(u_{B_1},v_{B_1}):=(1/x,x/y)=(0,-1)\\ &&B_3:
(u_{B_2},v_{B_2}):=(u_{B_1},\frac{v_{B_1}+1}{u_{B_1}}) =(0, -1/a)\\ &&B_4:
(u_{B_3},v_{B_3}):=(u_{B_2},\frac{v_{B_2}+1/a}{u_{B_2}}) =(0, -1/a^2)\\ &&C_i:
(x/y,y^i/x^{i-1}):=(0,0) \quad (i\geq 1)\\ &&D_1: (x^i/y^{i-1},y/x):=(0,0)
\quad
(i\geq 1)
\een

While the action of $(\vp_n)_*$  on the curves themselves is different from the
generic case, the action $(\vp_n)_*:\pic(X_n)\to \pic(X_{n+1})$  is the same as
(\ref{goract}). Hence the invariant effective class is the same as in the
generic
case.

\section{Conclusion}

In this paper, we have studied the space of initial conditions for a family of
linearisable mappings.
These mappings have been obtained either in studies of projective systems
or through limiting
procedures of discrete Painlev\'e equations. They are in general
nonautonomous and, except for
the projective cases, they possess unconfined singularities. Despite this
fact (which necessitates
in principle an infinite number of blow-ups) we were able to construct the
space of initial conditions
and the analytically stable sequence of surfaces.

We have also
shown that the degrees of iterated mappings can be calculated
by considering the action on the Picard groups of these
sequence of surfaces.
In the case of Riccati type equations, the degrees of which grow linearly
and in the autonomous case these belong to the ruled surface
case in \cite{df},
we have shown that \\
i) the corresponding
surfaces have fibrations such that the fibers are mapped to
the fibers of the next surface \\
ii) by using these fibrations we can
reduce such mapping to two projective mappings in cascade, i.e.
a projective mapping on the base space and
that on fibers, where the solution of the former one can be used
in the coefficients in the second one.

Many more mappings, which are integrable through linearisation or 
whose degree grows linearly, have been
derived
in the past few years. It would be interesting to examine them using the
tools we have
presented in this paper.

\end{document}